\documentclass[10pt,a4paper]{article}
\usepackage{graphicx,amsmath,color,psfrag}

\newcommand{\g}{\Gamma}
\newcommand{\z}[1]{z_{#1}}
\newcommand{\zb}[1]{\overline{z_{#1}}}
\newcommand{\pp}[1]{\partial_{#1}}
\newcommand{\hh}{h_{2}}
\newcommand{\la}{\langle}
\newcommand{\ra}{\rangle}

\begin{document}
\parindent 0mm
\parskip 6pt
\title{Correlation functions of twist operators applied to single self-avoiding loops}
\author{Adam Gamsa and John Cardy\\
Rudolf Peierls Centre for Theoretical Physics\\
         1 Keble Road, Oxford OX1 3NP,
         U.K.}
\date{June 2006}
\maketitle
\begin{abstract}
The O($n$) spin model in two dimensions may equivalently be
formulated as a loop model, and then mapped to a height model
which is conjectured to flow under the renormalization group to a
conformal field theory (CFT). At the critical point, the order $n$
terms in the partition function and correlation functions describe
single self-avoiding loops. We investigate the ensemble of these
self-avoiding loops using twist operators, which count loops which
wind non-trivially around them with a factor $-1$. These turn out
to have level two null states and hence their correlators satisfy
a set of partial differential equations. We show that
partly-connected parts of the four point function count the
expected number of loops which separate one pair of
points from the other pair, and find an explicit expression for
this. We argue that the differential equation satisfied by these
expectation values should have an interpretation in terms of a
stochastic(Schramm)-Loewner evolution (SLE$_{\kappa}$) process
with $\kappa=6$. The two point function in a simply connected
domain satisfies a closely related set of
equations. We solve these and hence calculate the expected number
of single loops which separate both points from the boundary.
\end{abstract}
\psfrag{a*}{(a)} \psfrag{b*}{(b)} \psfrag{c*}{(c)}
\psfrag{z1}{$(z_{1},\overline{z_{1}})$}
\psfrag{z2}{$(z_{2},\overline{z_{2}})$}
\psfrag{c1}{$(\overline{z_{1}},z_{1})$}
\psfrag{c2}{$(\overline{z_{2}},z_{2})$}
\psfrag{z3}{$(z_{3},\overline{z_{3}})$}
\psfrag{z4}{$(z_{4},\overline{z_{4}})$}
\psfrag{234}{$G_{z_{1}z_{2}z_{3}z_{4}|}$}
\psfrag{134}{$G_{z_{1}z_{3}z_{4}|z_{4}}$}
\psfrag{24}{$G_{z_{1}z_{2}z_{4}|z_{3}}$}
\psfrag{|2}{$G_{z_{1}z_{3}z_{4}|z_{2}}$}
\psfrag{23}{$G_{z_{1}z_{2}z_{3}|z_{4}}$}
\psfrag{14}{$G_{z_{1}z_{4}|z_{2}z_{3}}$}
\psfrag{13}{$G_{z_{1}z_{3}|z_{2}z_{4}}$}
\psfrag{12}{$G_{z_{1}z_{2}|z_{3}z_{4}}$}
\psfrag{1|}{$G_{z_{1}|z_{2}z_{3}z_{4}}$}
\psfrag{Gb}{$G_{1,z_{1}z_{2}|}$} \psfrag{Go}{$G_{1,z_{1}|z_{2}}$}
\psfrag{Gbn}{$G_{z_{1}z_{2}()}$} \psfrag{Gb1}{$G_{(z_{1})z_{2}}$}
\psfrag{Gb2}{$G_{z_{1}(z_{2})}$} \psfrag{Gbb}{$G_{(z_{1}z_{2})}$}
\psfrag{I}{Im} \psfrag{R}{Re}
\section{Introduction}
In a recent paper, Werner~\cite{WernerMeasure} has shown there
exists a measure on simple loops on any Riemann surface which
has the property of conformal restriction. This is to say that if
$D'\subset D$ are any two subdomains of the manifold, the measures
on loops in $D'$ obtained by (a) restriction of the measure on
loops in $D$ to those in $D'$, and (b) conformally mapping
$D\rightarrow D'$, are the same. Moreover this measure is unique
up to multiplication by a constant. We refer to these loops
throughout the paper as self-avoiding loops and to the mass of any
subset under Werner's measure as the $\mu$-mass.

One may also consider the set of self-avoiding polygons on some
regular lattice embedded in the manifold. The total number of such
polygons of length $l$ is known to grow as $\mu^{l}$ where $\mu$
is lattice dependent. The measure which weights each polygon with
a factor $x_{c}^{l}$ (where $x_{c}=\mu^{-1}$) has the restriction
property, and is commonly conjectured also to be conformally
invariant in the limit of vanishing lattice spacing, which means
that it should give a particular example of Werner's measure. In
this paper we assume this to be true, and hence conjecture
values of the $\mu$-mass of certain loop subsets, using
(non-rigorous) Coulomb gas and CFT methods applied to a
generalised model, the O$(n)$ model.

The O($n$) model encompasses a large group of physically
interesting models, including the Ising spin model, the percolation problem
and self-avoiding loops. The theory may be written as a
spin model with nearest neighbour interactions, or alternatively
in the loop gas picture, in which the states of the model are
non-intersecting loops constructed on the edges of the spin model.
Each loop is weighted with a factor $nx^{l}$, where $x$ is related
to the reduced coupling in the model and $l$ is the length of the
loop. In the Ising model, these loops are the cluster boundaries
of a spin model defined on the dual lattice. At the critical point of the
model, $x=x_{c}(n)$, it is conjectured to flow under the
renormalization group to a Gaussian free
field~\cite{NienhuisPhase,DotFat}, which is a well known example
of a conformal field theory (CFT).
One of us~\cite{CardyLinkingNumbers} showed the existence of operators
in this theory whose two point correlation functions count loops
around one of two points with a weight $n'$ rather than $n$, hence
giving information about the distribution of loops. The choice
$n'=-n$ is of particular interest; these operators are then called
twist operators and are at $(r,s)=(1,2)$ in the Kac
classification~\cite{bigyellow}. Hence, they have null states at
level two. The two point correlation function is then equivalent
to the expectation value of $(-1)^{N}$ in the loop ensemble, where
$N$ is the number of intersections of loops with a defect line, a
continuous curve connecting the two points. The twist operators
may therefore be thought of as a source and a sink for this defect
line.

\begin{figure}[h]
\centering
\includegraphics[width=0.6\textwidth]{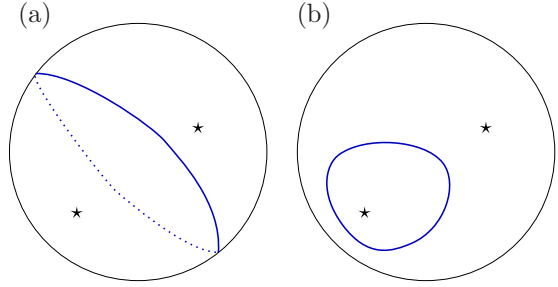}
\caption{Two examples of single loops winding around one of two operators for the theory on the Riemann sphere.
The stars mark the locations of the operators and the dashed line indicates that the curve
wraps around the back of the sphere.
Two point function is the sum of such loops, weighted by $x_{c}$ to the power of their length.
Notice that the sum contains loops with vanishingly small perimeter in the continuum limit.}\label{FigIntroTwoPoints}
\end{figure}

Self-avoiding loops are described by the loop gas picture of the
O($n$) model with $n\rightarrow 0$. The partition function and
correlation functions are determined (to order $n^{1}$) by graphs
with just a single loop. The two point correlation function of
twist operators therefore leads to an analytic expression for the
number of single loops which separate the locations of the
operators, weighted by $x_{c}^{l}$ as in figure~\ref{FigIntroTwoPoints}. This number is logarithmically
divergent in the continuum limit, however, due to the contribution
from vanishingly small loops around each of the points.

\begin{figure}[h]
\centering
\includegraphics[width=0.6\textwidth]{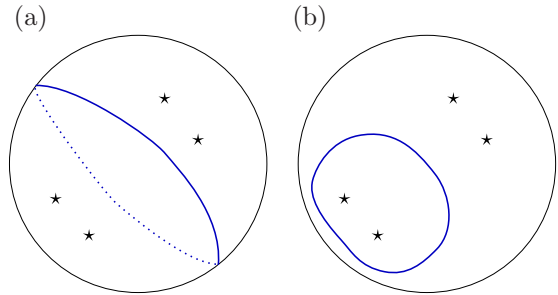}
\caption{Two examples of single loops winding around the locations
of two of four operators for the theory on the Riemann sphere. A
particular semi-connected four point correlation function counts
the number of such loops, weighted by $x_{c}$ to the power of
their length. Note that, in contrast to the two point correlation
function above, there are no vanishingly small loops contributing
to the sum. }\label{FigIntroFourPoints}
\end{figure}

In this paper, we consider the four point correlation function of
these twist operators on the Riemann sphere. In the loop gas
picture, we may think of each operator as again being a source (or
sink) of a defect line. Hence there is a defect line running
between each pair of points. We argue that the choice of paths for
the defect lines is unimportant and that they may run between any
two distinct pairs of points. The four point correlation function
can then be shown to be the expectation value of $(-1)^{N}$ where
$N$ is now the total number of crossings of loops with defect
lines. Particular semi-connected parts of this four point function
yield the expected numbers of weighted loops (the $\mu$-mass of
loops) which wind around two of the four locations of the
operators, as in figure~\ref{FigIntroFourPoints}. CFT may be used to derive the form of the four point
function, since the null states of the operators imply that the
correlation functions obey a set of Belavin, Polyakov and
Zamalodchikov (BPZ) type partial differential
equations~\cite{BPZ}. We solve these equations and hence find
analytic expressions for the $\mu$-mass of loops which separate
one pair of points from the other. For loops which separate
$(z_1,z_2)$ from $(z_3,z_4)$, for example, we obtain the
expression

\begin{align*}
\frac{-1}{24\pi}&\Big(\eta\,_{3}F_{2}(1,1,\frac{4}{3};2,\frac{5}{3};\eta)
+\overline{\eta}\,_{3}F_{2}(1,1,\frac{4}{3};2,\frac{5}{3};\overline{\eta})\Big)\\
&+\frac{2^{1/3}\pi}{3\sqrt{3}\Gamma(\frac{1}{6})^2\Gamma(\frac{4}{3})^2}
|\eta(1-\eta)|^{2/3}|\,_{2}F_{1}(\frac{2}{3},1,\frac{4}{3},\eta)|^{2}\,,
\end{align*}
where $\eta\equiv(z_{1}-z_{2})(z_{3}-z_{4})(z_{1}-z_{3})^{-1}(z_{2}-z_{4})^{-1}$ is the cross-ratio of the points.

These expected numbers of weighted loops are finite in the
continuum limit because there is no contribution from vanishingly
small loops, and are invariant under conformal transformations. We
show that the non-leading behaviour of these expressions as
$\eta\to0$ reveals the derivative of the central charge with
respect to $n$ at $n=0$. This is consistent with the
interpretation \cite{DoyonRivaCardy} of the stress tensor as the
spin-two component of the relative probability that a loop passes
between two points in the limit $z_{12}\to0$.

We use similar arguments for the O($n$) model in a simply
connected domain. Such a domain may always be mapped via a
conformal transformation to the upper half plane with the real
axis as the boundary. The two point function in this domain
satisfies the same partial differential equations as the four
point function in the bulk, with the locations of the two
operators and the reflections of these points in the real axis
being the locations of the four operators in the bulk theory. We
show that the connected two point function of twist operators in
the $n\rightarrow 0$ limit counts the mass of loops around the
locations of both operators, and find the explicit expression

\begin{align*}
-&\frac{1}{12\pi}\ln(\eta(1-\eta))-\frac{1}{12\pi}\eta\,_{3}F_{2}(1,1,\frac{4}{3};2,\frac{5}{3};\eta)\\
&+\frac{\Gamma(2/3)^2}{6\pi\Gamma(4/3)}(-\eta(1-\eta))^{\frac{1}{3}}\,_{2}
F_{1}(\frac{2}{3},1;\frac{4}{3};\eta)\,,
\end{align*}
where $\eta$ is now the cross ratio of the two points together
with their conformal images in the boundary.

The layout of this paper is as follows. In the next section, we recall the arguments leading to the
Coulomb gas picture of the O($n$) model. In section~\ref{SecTwistOperators} we discuss twist operators
in the O($n$) model and derive the form of the two point correlation function of these operators.
The small
$n$ limit of the theory is explored and the correlation function in this limit is shown to depend (to order $n^{1}$) only
on the configurations of a single loop. In section~\ref{Sec4ptFnCFT} we discuss the
four point correlation function of twist operators. Because the twist operators have null states at level
two, this correlation function satisfies a set of partial differential equations, which we solve analytically.
The interpretation of the four point function in terms of a pair of defect lines is explained, followed by the example
of the four point function of twist operators as the four point function of spin operators in the Ising model.
Section~\ref{Sec4ptInterpretation} is devoted to the small $n$
limit of the four point function and contains a derivation of the mass
of loops winding around two of the four points.
We then explain how these numbers, as functions of the positions of the four points, hold a key to measuring
the effective central charge of the O($n$) model as $n\rightarrow 0$.
In section~\ref{Sec2ptBCFT} we apply the theory of twist operators
to the O($n$) model in a simply connected domain and show that the mass of loops around two points
in such a domain is invariant under conformal transformations and finite in the limit of vanishing lattice spacing.
Finally, in section~\ref{SecStochasticProcess} we show that the mass of loops around two of four points
on the Riemann sphere can be thought of in terms of an SLE$_{6}$ process.
\section{The Coulomb Gas}
Let us start with the following partition function for the O($n$)
model \cite{NienhuisPhase}:
\begin{equation}\label{EqZON}
Z_{O(n)}=\textrm{Tr}\prod_{(ij)}{(1+x\textbf{s}(r_{i})\cdot
\textbf{s}(r_{j}))}\,.
\end{equation}
The $\textbf{s}(r_{i})$ are $n$-component spins on a lattice
$\{r_{i}\}$ and the product is over pairs of nearest
neighbours.
The product in the partition function can be expanded
into a sum of $2^N$ terms, where $N$ is the number of nearest neighbours.
Each term can be represented by a graph of open and closed loops in the following way:
the neighbouring sites, $i$ and $j$, are joined together
by a line if the $x\textbf{s}(r_{i})\cdot \textbf{s}(r_{j})$ term
was chosen in the expansion of equation~(\ref{EqZON}), or left disconnected if the number $1$ was chosen instead.
The trace over spins may then be done for each graph individually.
Since $\textrm{Tr}\,\textbf{s}(r_{i})^{\textrm{odd}}=0$ because of the symmetry
under the transformation $\textbf{s}(r_{i})\rightarrow-\textbf{s}(r_{i})$,
all graphs containing an odd number of $\textbf{s}(r_{i})$
make no contribution to the partition function.
These are the graphs with open loops, hence the partition function
will only contain contributions from graphs with closed loops.
Let us consider the honeycomb lattice for simplicity, then
there may be 0,1,2 or 3 powers of a given $\textbf{s}(r_{i})$.
The trace over spins of $\textbf{s}(r_{i})^{0}$ will contribute a factor of 1,
whereas, from $\textrm{Tr}\, s_{a}(r)s_{b}(r)=\delta_{ab}$, instances
of $\textbf{s}(r_{i})^{2}$ lead to a factor of $n$ for each closed loop
in the graph. The partition function is therefore equivalent to
\begin{equation}\label{EqLoopPartition}
Z_{O(n)}=\sum_{G}{x^{l}n^{v}}\,,
\end{equation}
where $l$ is the total length of all the loops in a given graph, $v$ is the number of
closed loops and the trace is now over all graphs of closed, non-intersecting
loops. This form of the partition function may be used for general
values of $n$, whereas equation~(\ref{EqZON}) is applicable only to positive, integer $n$.
There exists a critical point in the theory, at $x=x_{c}$,
where the mean loop
length diverges and the model is supposed to become conformally invariant. In order
to formulate a field theory, the non-local factors of $n$ must be made local.
This can be done by assigning orientations to the loops (either clockwise or anticlockwise)
and inserting local factors of $e^{i\chi}$ at each vertex where
the curve turns to the right and $e^{-i\chi}$ where the curve turns to the left.
Summing over the two possible orientations for each loop
leads to a contribution of $e^{6i\chi}+e^{-6i\chi}$
from each closed loop on the honeycomb lattice, where closed loops
have a difference in the number of left and right turns of six with the
sign depending on the orientation of the loop.
In order to obtain the desired factor of $n$ for each closed loop,
$\chi$ must therefore be chosen such that
\begin{equation}\label{EqChiN}
e^{6i\chi}+e^{-6i\chi}=2\cos6\chi=n\,.
\end{equation}
This may now be transformed into a height model. The height
variables are taken to be integer multiples of $\pi$ and are
assigned to sites on the dual lattice, such that the loops are
contours of the landscape. Crossing a loop running from left to
right leads to a decrease in the height of $\pi$, whilst crossing
a loop running from right to left leads to an increase in the
height of $\pi$. A given configuration of the heights corresponds
to a unique graph of orientated loops. The assumption of the
Coulomb gas is that this height model flows under the
renormalisation group (RG) into a free field theory with action
$$
S[\, h(\textbf{r})]=\frac{g}{4\pi}\int{(\partial h(\textbf{r}))^2 d^{2}\textbf{r}}\,,
$$
for some $g(n)$. There is an additional subtlety regarding topology,
which may be seen by considering the model defined on the cylinder.
On a cylinder of circumference~$L$, the loops wrapping around
the circumference are counted incorrectly (for~$n\neq 2$).
Loops wrapping around have the same number of left turns as right turns.
These loops are therefore not counted correctly.
The situation is remedied by placing a vertex operator (also known as an electric charge) $e^{i6\chi h/\pi}$
at one end of the cylinder and an operator $e^{-i6\chi h/\pi}$ at the other end.
Let these points be $\pm w/2$.
If there is a single
loop wrapping around the cylinder, then there will be a height difference of $\pi$ between the ends,
the sign of which depends on the orientation of the loop. Summing over the two possibilities for the
orientation then leads to the required contribution of $e^{i6\chi}+e^{-i6\chi}$.
These charges at the ends of the cylinder lead to a modification of the partition function to
\begin{align*}
Z_{\textrm{Coulomb Gas}}&=Z_{\textrm{ffc}}
\langle e^{-i\frac{6\chi}{\pi}
h(-w/2)}e^{i\frac{6\chi}{\pi}h(w/2)}\rangle_{\textrm{ffc}}\\
&=Z_{\textrm{ffc}}
\Big| \frac{L^2}{2\pi^2}(\cosh(\frac{2\pi w}{L})-1)\Big|^{(6\chi)^2/2\pi^2 g}\\
&=Z_{\textrm{ffc}} \Big|\frac{L^2}{4\pi^2}e^{2\pi w/L}\Big|^{(6\chi)^2/2\pi^2 g}\,,
\end{align*}
where the abbreviation ffc signifies that the above expectation values
and partition functions are for the free field theory on the cylinder
and where we have used $w\gg L$.
The free energy per unit length on the cylinder for this Coulomb gas (CG) partition function
may be calculated as
$$
\frac{F_{CG}}{w}=-\frac{\ln(Z_{CG})}{w}=-\frac{\pi c}{6L}\,.
$$
This is the form of the free energy on the cylinder obtained via the Schwartzian derivative from
a theory on the plane with central charge $c$~\cite{bigyellow} given by
\begin{equation}
c=1-\frac{6}{g}\Big(\frac{6\chi}{\pi}\Big)^{2}\,,
\end{equation}
where the first term comes from the known behaviour of $Z_{ffc}$~\cite{bigyellow}.
All that remains is to fix the value of $g$, which may be obtained
from the following argument: adding a term $-\lambda \int
\cos(2h)d^{2}r$ to the action should not affect the critical
behaviour, since the height variables were defined to be integer
multiples of $\pi$ and $\cos(2m\pi)=1$ for all integer $m$. Hence,
this term should be marginal under renormalisation group flow and
therefore must have scaling dimension~$2$. This determines~\cite{Kondev}
\begin{equation}\label{EqGChi}
g=1-6\chi/\pi\,.
\end{equation}
\section{Twist operators}\label{SecTwistOperators}
\begin{figure}[t]
\centering
\includegraphics[width=0.7\textwidth]{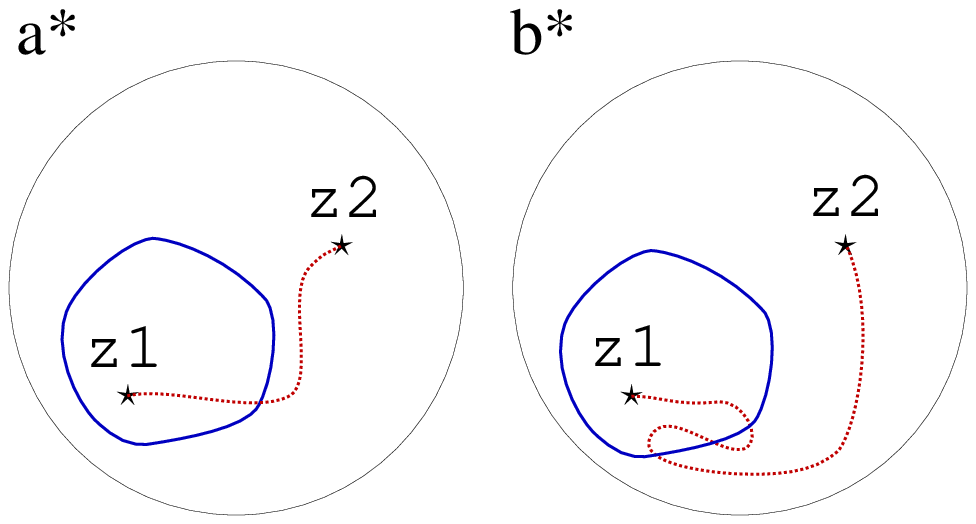}
\caption{A graph on the Riemann sphere belonging to $\{G_{\textrm{odd}}\}$ with one loop (solid line)
and two choices of path for the defect line (dashed line).
In (a) $N_{12}=1$ so that $(-1)^{N_{12}}=-1$.
In (b) $N_{12}=3$ so that $(-1)^{N_{12}}=-1$ also.
Any choice of path for the defect line will lead to $(-1)^{N_{12}}=-1$ since $N_{12}$ is always odd.
}\label{FigDefectLine}
\end{figure}

In the loop (Coulomb) gas picture of the $O(n)$ model, loops wrapping around the cylinder
may be counted with a weight $n'$ different from $n$
by placing additional charges $e^{\pm 6ih(\chi'-\chi)/\pi}$ at the ends of
the cylinder, with $\chi'$ chosen according to equation~(\ref{EqChiN}).
The scaling dimension of these operators may be obtained from their two
point correlation function in the loop gas ensemble. Placing the charges at $\pm s/2$,
$$
\lim_{w\gg L}\frac{\langle e^{-i\frac{6\chi}{\pi}
h(-w/2)}e^{-i\frac{6(\chi'-\chi)}{\pi}
\frac{h}{\pi}(-s/2)}e^{i\frac{6(\chi'-\chi)}{\pi}
h(s/2)}e^{i\frac{6\chi}{\pi}h(w/2)}\rangle_{\textrm{free-field, cyl.}}}
{\langle e^{-i\frac{6\chi}{\pi}
h(-w/2)}e^{i\frac{6\chi}{\pi}h(w/2)}\rangle_{\textrm{free-field, cyl.}}}\,.
$$
These free field expectation values may be calculated explicitly since they are Gaussian moments.
Then, in the limit $s\gg L$, the two point function is seen to be
\begin{equation*}
\Big|\frac{L^2}{4\pi^2}e^{2\pi w/L}\Big|^{-6^2(\chi'^2-\chi^2)/2\pi^2 g}\,.
\end{equation*}
This expression is of the general form of a two point function of operators on the cylinder
with scaling dimension~\cite{CardyOpsOnCyl}
\begin{equation}\label{EqXFnOfChi}
x(n,n')=6^{2}(\chi'^2-\chi^2)/2\pi^2 g\,.
\end{equation}
A particular choice of interest is $n'=-n$, for which the scaling dimension is $x(n,-n)=3/2g-1$;
this is the scaling dimension of $\phi_{1,2}$ operators with level two null states in the Kac classification:
$$
x_{r,s}=\frac{(rg-s)^2-(g-1)^2}{2g}\,.
$$
These correspond, in string theory language, to operators which
insert orbifold points corresponding to the global symmetry $\bf
s\to\bf -s$ of the hamiltonian and are known as twist operators.
They will be the focus of this paper. Their correlation functions
may be considered in geometries other than the cylinder. For
example, on the Riemann sphere, the two point function of such
operators is fixed by scale invariance to be
\begin{equation}\label{Eq2ptCFT}
\langle \phi(z_{1},\overline{z_{1}})\phi(z_{2},\overline{z_{2}})\rangle_{\textrm{CFT}}=
\Big|\frac{z_{1}-z_{2}}{a}\Big|^{\, 2-3/g(n)}\,,
\end{equation}
where we have inserted explicit factors of the lattice spacing, $a$,
so as to make $\phi$ dimensionless.
The two point correlation function in the loop gas may also be calculated on
the sphere using the ensemble of equation~(\ref{EqLoopPartition}).
Each loop separating the points $(\z 1,\zb 1)$ and $(\z 2,\zb 2)$ will be counted with weight $-n$ rather than
$n$, hence graphs in $G$ will be weighted with an additional factor of $(-1)$ to the power of the number
of loops separating the two points.
Graphs in $G$ with an odd number of loops separating point $(\z 1,\zb 1)$ from point $(\z 2,\zb 2)$
will be weighted by an additional factor of $(-1)^{\textrm{odd}}=-1$, whilst those with an even number of loops will be
unaffected, since $(-1)^{\textrm{even}}=1$. These two different types of graphs can be separated in
the sum over graphs as follows
\begin{equation}\label{Eq2ptLoops}
\langle \phi(z_{1},\overline{z_{1}})\phi(z_{2},\overline{z_{2}})\rangle_{\textrm{loop gas}}=
\Big[\sum_{G_{\textrm{odd}}(\z 1,\z 2)}{(-1)x_{c}^{\textrm{l}}n^{\textrm{v}}}
+\sum_{G_{\textrm{even}}(\z 1,\z 2)}{x_{c}^{\textrm{l}}n^{\textrm{v}}}\Big]\Big/Z_{\textrm{loop gas}}\,,
\end{equation}
where $G_{\textrm{odd}}(\z 1,\z 2)$ is the set of graphs of closed, non-intersecting loops with an odd number of
loops separating point $(\z 1,\zb 1)$ and $(\z 2,\zb 2)$ and the $G_{\textrm{even}}(\z 1,\z 2)$
is the set with an even number of loops separating the two points.
This two point correlation function has a natural interpretation in terms of a defect line
joining points $(\z 1,\zb 1)$ and $(\z 2,\zb 2)$.
If $N_{12}$ is the number of times loops cross the defect line in a given graph, then
equation~(\ref{Eq2ptLoops}) may be rewritten as
\begin{align}
\la\phi(z_{1},\overline{z_{1}})\phi(z_{2},\overline{z_{2}})\rangle_{\textrm{loop gas}}&=
\sum_{G}
{(-1)^{N_{12}}x_{c}^{\textrm{l}}n^{\textrm{m}}}\Big/Z_{\textrm{loop gas}}\nonumber\\
&=\la (-1)^{N_{12}}\ra_{\textrm{loop gas}}\label{EqLoopsDefect}\,.
\end{align}
The defect line can take any path between the two points because, for a given loop configuration, $(-1)^{N_{12}}$
will be the same for all paths. This can be best understood with
reference to figure~\ref{FigDefectLine}.
Combining equations~(\ref{Eq2ptCFT})~and~(\ref{EqLoopsDefect}), we find
\begin{equation}\label{Eq2ptInterpretation}
\la (-1)^{N_{12}}\ra_{\textrm{loop gas}}=
\Big|\frac{z_{1}-z_{2}}{a}\Big|^{\, 2-3/g(n)}\,.
\end{equation}
\subsection{The small $n$ limit}\label{Sec2ptsmalln}
Self-avoiding loops correspond to $g=3/2$, which is the dilute phase of the $n\rightarrow 0$ limit
of the $O(n)$ model. The results of section~\ref{SecTwistOperators} are summarized by
equation~(\ref{Eq2ptInterpretation}), which for $n\ll 1$ may be expanded in powers of $n$.
The order $n^1$ term in the loop gas expansion (the left hand side of the equation)
comes from graphs with a single loop.
By equating this with the order $n^1$ term of the right hand side,
we shall demonstrate a property of the sum over graphs with a single loop.

\begin{figure}[h]
\centering
\includegraphics[width=0.7\textwidth]{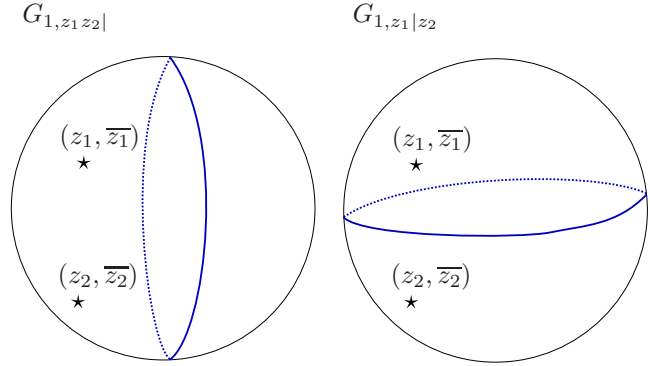}
\caption{An example of a graph belonging to $G_{1,\z 1\z 2|}$ and
one belonging to $G_{1,\z 1|\z 2}$ for a choice of $(\z 1,\zb 1)$ and $(\z 2,\zb 2)$
on the Riemann sphere.
Notice that a loop in $G_{1,\z 1|\z 2}$ can be thought of as being
either around $(\z 1,\zb 1)$ or around $(\z 2,\zb 2)$.
Similarly, a loop in $G_{1,\z 1\z 2|}$ can be thought of as being around both points or neither.
}\label{Fig2pt1loop}
\end{figure}

The left hand side of equation~(\ref{Eq2ptInterpretation}) for $n\ll 1$ becomes
\begin{align}\label{Eq2ptSmallNLHS}
\Big(\sum_{G_0}(-1)^{N_{12}}x_{c}^{l}n^{0}+\sum_{G_1}(-1)^{N_{12}}x_{c}^{l}n^{1}\Big)
\Big(\sum_{G_0}x_{c}^{l}n^{0}+\sum_{G_1}x_{c}^{l}n^{1}\Big)^{-1}+O(n^2)\,,
\end{align}
where $G_{0}$ is the set of all graphs with no loops and $G_{1}$ is the set
of all graphs with one loop, hence their respective powers of $n^{0}$ and $n^1$.
$G_{0}$ contains a single graph with no edges, hence the first term in both
brackets is equal to $1$. The set $G_{1}$ contains graphs with only a single loop. This set
is composed of two subsets with no overlap, the set $G_{1,\z 1|\z 2}$ with a single loop
separating the point $(\z 1,\zb 1)$ from $(\z 2,\zb 2)$ and the set $G_{1,\z 1\z 2|}$ which
contains no loops separating the points.
It should be noted that a single loop surrounding the point $(\z 1,\zb 1)$ is classified in the same group
as a loop around $(\z 2,\zb 2)$ since on the Riemann sphere one
can be continuously deformed into the other.
Both have an odd number of intersections with a defect line between the two points,
hence $(-1)^{N_{12}}=-1$. For all graphs belonging to $G_{1,\z 1\z 2|}$, it may be seen that $(-1)^{N_{12}}=1$.
This is shown in figure~\ref{Fig2pt1loop}.

Equation~(\ref{Eq2ptSmallNLHS}) can therefore be rewritten as
\begin{align}
\Big(1+&\sum_{G_{1,\z 1|\z 2}}(-1)x_{c}^{l}n^{1}+\sum_{G_{1,\z 1\z 2|}}x_{c}^{l}n^{1}\Big)
\Big(1+\sum_{G_{1,\z 1|\z 2}}x_{c}^{l}n^{1}+\sum_{G_{1,\z 1\z 2|}}x_{c}^{l}n^{1}\Big)^{-1}+O(n^2)\nonumber\\
&=1-2n\sum_{G_{1,\z 1|\z 2}}x_{c}^{l}+O(n^2)\label{Eq2ptSmallNLoopExpansion}\,.
\end{align}
Equations~(\ref{EqXFnOfChi}),~(\ref{EqChiN})~and~(\ref{EqGChi}) may be combined and then expanded in powers of $n$ to find
the following form for the scaling dimension of the twist operators in the small $n$ limit:
$$
x(n)=-1+3/2g(n)=n/3\pi+O(n^2)\,.
$$
Therefore, the expansion of the right hand side of equation~(\ref{Eq2ptInterpretation}) is
\begin{equation}\label{Eq2ptCFTSmallN}
\Big|\frac{\z 1-\z 2}{a}\Big|^{-2n/3\pi}+O(n^2)=1-\frac{2n}{3\pi}\ln{\Big|\frac{\z 1-\z 2}{a}\Big|}+O(n^2)
\end{equation}
Hence, from comparing the coefficients of $n^1$ in equations~(\ref{Eq2ptSmallNLoopExpansion})~and~(\ref{Eq2ptCFTSmallN}),
we may conclude that
\begin{equation}\label{Eq2ptWeightResult}
\sum_{G_{1,\z 1|\z 2}}x_{c}^{l}=\frac{1}{3\pi}\ln{\Big|\frac{\z 1-\z 2}{a}\Big|}\,.
\end{equation}
In words, for a given pair of points $(\z 1,\zb 1)$ and $(\z 2,\zb
2)$, the number of weighted loops which separate the two points,
weighted by $x_{c}^{l}$, diverges as $(1/3\pi)\ln{|1/a|}$ in the
limit of the lattice spacing tending to zero. The cause of this is
the diverging contribution from vanishingly small loops in the
continuum limit. The factor $x_{c}^{l}$ may be thought of as the
measure, hence this weighted number of loops is equivalent to the
expected number of loops around one point. It may be noted that
the coefficient $1/3\pi$ differs from that seen on the annulus
with shrinking internal radius. In the case of the annulus with
vanishingly small modulus~\cite{JCAnnulus} a coefficient of
$1/6\pi$ is seen, because there are only diverging contributions
from small loops around one point, as opposed to the plane where
small loops around both points contribute.
\section{The four point correlation function of twist operators from conformal field theory}\label{Sec4ptFnCFT}
Twist operators were introduced in section~\ref{SecTwistOperators} as the operators responsible
for counting loops with weight $-n$ rather than $n$ in the loop gas picture of the $O(n)$ model.
Their scaling dimension was calculated
and it was seen that it corresponds to the scaling dimension of operators with null states at level two.
Conformal field theory may be used to derive a set of partial differential equations satisfied
by the correlation functions of such null state operators. In the case of the four point function,
these differential equations may be solved analytically, as will be seen in this section.
To make contact with the language of Schramm (stochastic)-Loewner Evolution (SLE), $g$ is dropped
in favour of the parameter $\kappa$ used in SLE~\cite{CardyRev}. The two are related by the formula $\kappa=4/g$.

It was shown in section~\ref{SecTwistOperators} that the twist operators have a null state at level two~\cite{bigyellow}.
This null state is itself a highest weight state; it is annihilated by all raising operators, $L_{n}$ with $n>0$.
This is not the same as saying that the null state decouples from all other states in the theory,
as it does in unitary theories.
That said, it can be shown by a modular invariance argument that physical results are seen only if the
null state decouples on the torus.
This is because, barring unforeseen cancellations, the nondecoupling of the null states would lead to a density
of states at high energies corresponding to $c=1$ rather than $c<1$~\cite{JCNucl}.
We shall therefore set the null state to zero in this theory and
use the resulting partial differential equations for the correlation functions of the twist operators.
In the complex plane, the four point function
satisfies the following set of partial differential equations (for~$j=1,2,3,4$)~\cite{BPZ}:
\begin{align}
&\Big[\pp {\z j}^2-\frac{\kappa}{4}\sum_{i\neq j}{\frac{h_{2}}{(\z
i-\z j)^2}-\frac{\pp{\z i}}{\z i-\z j}}\Big]\la \phi(\z 1,\zb 1)\phi(\z 2,\zb 2)
\phi(\z 3,\zb 3)\phi(\z 4,\zb 4)\ra=0\label{Eq4ptBPZHolomorphic}\\
&\Big[\pp {\overline{\z j}}^2-\frac{\kappa}{4}\sum_{i\neq j}{\frac{\overline{h_{2}}}{(\zb
i-\zb j)^2}-\frac{\pp{\overline{\z i}}}{\zb i-\zb j}}\Big]\la \phi(\z 1,\zb 1)\phi(\z 2,\zb 2)
\phi(\z 3,\zb 3)\phi(\z 4,\zb 4)\ra=0.\label{Eq4ptBPZAntiholomorphic}
\end{align}
In terms of $z_{ij}\equiv\z i-\z j$, the cross ratios are defined as
\begin{align*}
&\eta\equiv\frac{z_{12}z_{34}}{z_{13}z_{24}}
&\overline{\eta}\equiv\frac{\overline{z_{12}}\,\overline{z_{34}}}{\overline{z_{13}}\,\overline{z_{24}}}\,.
\end{align*}
These cross ratios are invariant under global conformal transformations.
The partial differential equations above have the solutions
\begin{equation}\label{Eq4ptCFTResult}
\la\phi(z_{1},\overline{z_{1}})\phi(z_{2},\overline{z_{2}})\phi(z_{3},\overline{z_{3}})\phi(z_{4},\overline{z_{4}})
\ra_{\textrm{CFT}}=\Big|\frac{z_{13}z_{24}a^2}{z_{12}z_{34}z_{23}z_{14}}\Big|^{4\hh}A(\kappa)\xi(\eta,\overline{\eta},\kappa)\,,
\end{equation}
where $G(\eta,\overline{\eta},\kappa)$ is
\begin{align*}
\xi(\eta,\overline{\eta},\kappa)&=
|_{2}F_{1}(1-\frac{\kappa}{4},2-\frac{3\kappa}{4};2-\frac{\kappa}{2};\eta)|^{2}\\
+&B(\kappa)|\eta(1-\eta)|^{2h_{3}}|_{2}F_{1}(\frac{\kappa}{4},\frac{3\kappa}{4}-1;\frac{\kappa}{2};\eta)|^{2}\,,
\end{align*}
$h_{2}=3\kappa/16-1/2$ and $h_{3}=\kappa/2-1$ is the value of the scaling dimension
of an operator with a null state at level three.
The form of $B(\kappa)$ can be derived by requiring consistency under the permutation of the labels
$(z_{i},\overline{\z i})$ and is found to be~\cite{gradryz,AbSt}
\begin{align*}
B(\kappa)&=\frac{[\g(1-\frac{\kappa}{4})^{2}\g(\frac{\kappa}{4})^{2}-\g(2-\frac{\kappa}{2})^{2}\g(\frac{\kappa}{2}-1)^{2}]
\g(\frac{3\kappa}{4}-1)^{2}}{\g(\frac{\kappa}{2})^2\g(\frac{\kappa}{2}-1)^2\g(1-\frac{\kappa}{4})^2}\,.
\end{align*}
The form of $A(\kappa)$ is dependent on the normalisation of the two point function since, as
$\eta\rightarrow 0$, the four point function becomes
$$
\la\phi(z_{1},\overline{z_{1}})\phi(z_{2},\overline{z_{2}})\phi(z_{3},\overline{z_{3}})\phi(z_{4}
,\overline{z_{4}})\ra_{\textrm{CFT}}
\rightarrow A(\kappa)\big|\frac{z_{12}z_{34}}{a^2}\big|^{-4h_{2}}\,.
$$
\subsection{Interpretation of the four point correlation function}\label{Sec4ptInterpretation}
\begin{figure}[t]
\centering
\includegraphics[width=1\textwidth]{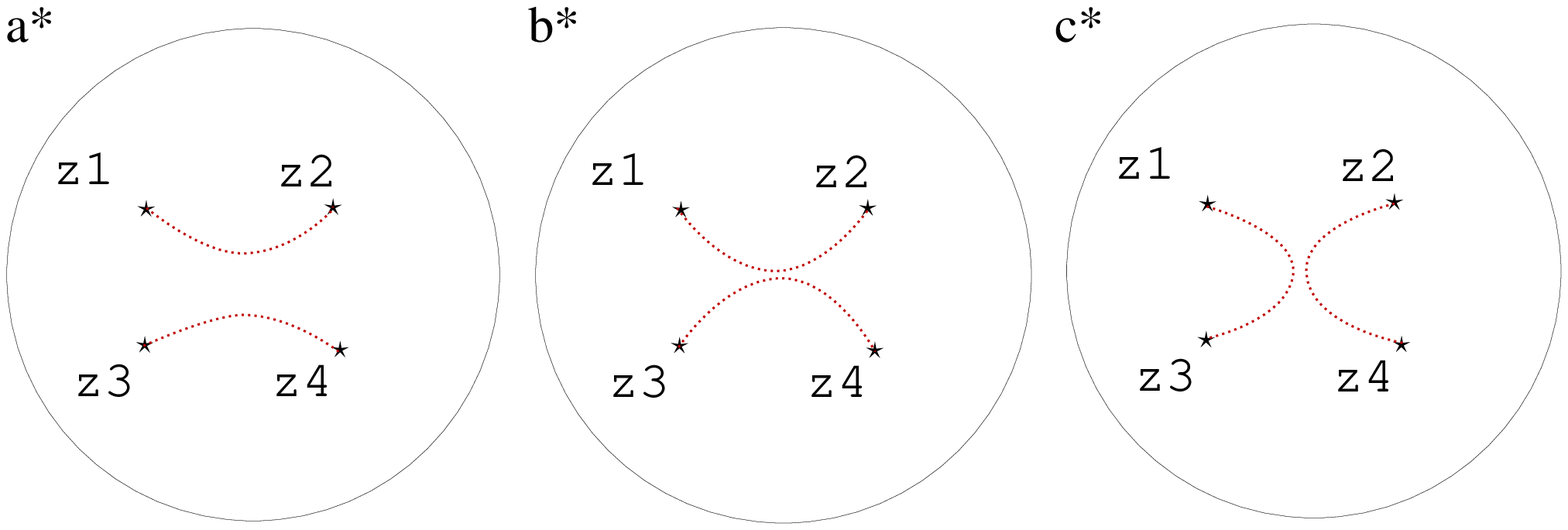}
\caption{(a) shows the choice of defect lines described in the text.
As the defect lines may take any path between their endpoints, they
may be re-routed to pass infinitesimally close to each other, as in (b).
Since a loop crossing a defect line leads to an additional weight $(-1)$ in the loop gas partition function,
a pair of defect lines leads to a factor $(-1)^2=1$ and hence makes no contribution to the
weight of any graph. Hence, the defect lines shown in (b) are equivalent to the choice in (c).
A similar construction may be used to show the equivalence to a pair of defect lines
drawn from $(\z 1,\zb 1)$ to $(\z 2,\zb 2)$ and from $(\z 3,\zb 3)$ to $(\z 4,\zb 4)$.
Note that this feature is a result of the choice $n'=-n$ in section~\ref{SecTwistOperators}.
}\label{FigDeformDefects}
\end{figure}

Recall from section~\ref{SecTwistOperators} that the correlation function of two $\phi$ operators
has an interpretation in terms of a defect line in the loop gas ensemble.
The operators are the source and sink
of the defect line, which can take any path between the two points.
Each graph in the sum is weighted by an additional factor
$(-1)^{N_{12}}$ where $N_{12}$ is the number of intersections of loops in that graph
with the defect line. There is a similar interpretation of the
four point function in the loop gas ensemble, but now there are two defect lines.
The four points are split into two pairs and defect lines run between the points in each pair.
For the choice of pairing ($\z 1,\zb 1$) with ($\z 2,\zb 2$)
and ($\z 3,\zb 3$) with ($\z 4,\zb 4$), the four point function
can be seen to correspond to the following expectation value in the loop gas picture
\begin{equation}\label{Eq4ptLoopResult}
\la\phi(z_{1},\overline{z_{1}})\phi(z_{2},\overline{z_{2}})\phi(z_{3},\overline{z_{3}})\phi(z_{4}
,\overline{z_{4}})\ra_{\textrm{loop gas}}=\la (-1)^{N_{12}} (-1)^{N_{34}}\ra_{\textrm{loop gas}}\,,
\end{equation}
where $N_{12}$ is the number of crossings of loops across a defect line from $(z_{1},\overline{z_{1}})$
to $(z_{2},\overline{z_{2}})$ and $N_{34}$ is similarly defined.
There are three ways in which the pairs may be chosen, but each choice leads to the same set of weights
for the graphs in the partition function.
The reason for this may be seen in figure~\ref{FigDeformDefects}.
Thus, the four point function may also be considered to be the expectation value in the loop gas ensemble
of $(-1)^{N_{13}} (-1)^{N_{24}}$ or $(-1)^{N_{14}} (-1)^{N_{23}}$.
Comparing equations~(\ref{Eq4ptCFTResult})~and~(\ref{Eq4ptLoopResult}) leads to the main result of this section
\begin{equation}\label{Eq4ptInterpretation}
\la (-1)^{N_{12}} (-1)^{N_{34}}\ra_{\textrm{loop gas}}=
\Big|\frac{z_{13}z_{24}a^2}{z_{12}z_{34}z_{23}z_{14}}\Big|^{4\hh}A(\kappa)\xi(\eta,\overline{\eta},\kappa)\,,
\end{equation}
where $\xi(\eta,\overline{\eta},\kappa)$ is defined below equation~(\ref{Eq4ptCFTResult}) in section~\ref{Sec4ptFnCFT}.

\subsection{Example: the Ising model}
Boundaries between Ising spin clusters correspond to the loops of the $O(n)$ model at $n=1$,
or $\kappa=3$.
The twist operators in this case are equivalent to magnetisation operators because
the parity of the number of loops separating two spins depends on whether they are
parallel or anti-parallel.
That is to say that
\begin{align*}
\la (-1)^{N_{12}}\ra_{\textrm{loop gas}}&=\Big|\frac{a}{z_{1}-z_{2}}\Big|^{4h_{2}}\\
&=\la s(\z 1,\zb 1)s(\z 2,\zb 2)\ra_{\textrm{Ising}}\,.
\end{align*}
Similarly, the four point correlation function of the twist operators is the four point
function of the magnetisation operators.
The following function is the result of substituting $\kappa=3$ into the results of section~\ref{Sec4ptFnCFT}
\begin{align}
\la\phi(z_{1},\overline{z_{1}})\phi(z_{2},\overline{z_{2}})\phi(z_{3},\overline{z_{3}})\phi(z_{4}
,\overline{z_{4}})\ra &=\frac{A(3)}{2}\Big|\frac{z_{13}z_{24}a^2}{z_{12}z_{34}z_{23}z_{14}}\Big|^{1/4}
(\, |1+\sqrt{\eta}\,|+|1-\sqrt{\eta}\,|\,)\label{EqIsing4ptResult}\,.
\end{align}
Equation~(\ref{EqIsing4ptResult}) is the well known four point
correlation of spin operators in the Ising model at criticality~\cite{BPZ}.
\section{The small $n$ limit of the four point function}\label{Sec4ptSmallN}
In section~\ref{SecTwistOperators}, we examined the small $n$ limit of the two point function of
twist operators, equation~(\ref{Eq2ptInterpretation}).
We found an exact expression for the $\mu$-mass of loops surrounding one of
the two points. This result is summarised by equation~(\ref{Eq2ptWeightResult}).
A similar expansion to order $n^1$
of the four point correlation function (equation~\ref{Eq4ptInterpretation}) yields
information about the configurations of a single self-avoiding loop around four points.

From the equations in section~\ref{Sec4ptFnCFT}, the following small $n$ expansions may be deduced
\begin{align*}
\kappa&=\frac{8}{3}+\frac{8}{9\pi}n+O(n^{2})\\
\hh&=\frac{n}{6\pi}+O(n^{2})\\
h_{3}&=\frac{1}{3}+\frac{4}{9\pi}n+O(n^{2})\\
B&=\frac{8 (2)^{1/3}\pi}{3\sqrt{3}\Gamma(\frac{1}{6})^2\Gamma(\frac{4}{3})^2} n
+O(n^{2})\,.
\end{align*}

The expansion of $A(n)$ around $n=0$ is
$A(n)=1+\varrho n+O(n^2)$ where $\varrho$ is a constant independent of $n$,
since the correlation function should tend to $1$ as $n\rightarrow 0$.
The Taylor series expansion of the right hand side of equation~(\ref{Eq4ptInterpretation}) is then
\begin{align}
\la\phi&(z_{1},\overline{z_{1}})\phi(z_{2},\overline{z_{2}})\phi(z_{3},\overline{z_{3}})\phi(z_{4}
,\overline{z_{4}})\ra_{\textrm{CFT}}=\nonumber\\
&1+n\Big[\varrho-\frac{2}{3\pi}\ln|\eta\, z_{23}z_{14}a^{-2}|-\frac{1}{3\pi}\Big(\eta\,_{3}F_{2}(1,1,\frac{4}{3};2,\frac{5}{3};\eta)
+\overline{\eta}\,_{3}F_{2}(1,1,\frac{4}{3};2,\frac{5}{3};\overline{\eta})\Big)\nonumber\\
&\quad+\frac{8 (2)^{1/3}\pi}{3\sqrt{3}\Gamma(\frac{1}{6})^2\Gamma(\frac{4}{3})^2}
|\eta(1-\eta)|^{2/3}|\,_{2}F_{1}(\frac{2}{3},1;\frac{4}{3};\eta)|^{2}\Big]+O(n^2)\label{Eq4ptSmallNCFT}\,.
\end{align}
The left hand side of equation~(\ref{Eq4ptInterpretation})
may also be expanded in powers of $n$. The graphs up to order $n^1$ in $G$ may be split
into the graph with no loops $(n^0)$ and eight distinct sets of graphs with a single loop,
as will be shown in the next section.
\subsection{The configurations of a single loop around four points}\label{Sec1loop4points}
\begin{figure}[!p]
\centering
\includegraphics[width=1\textwidth]{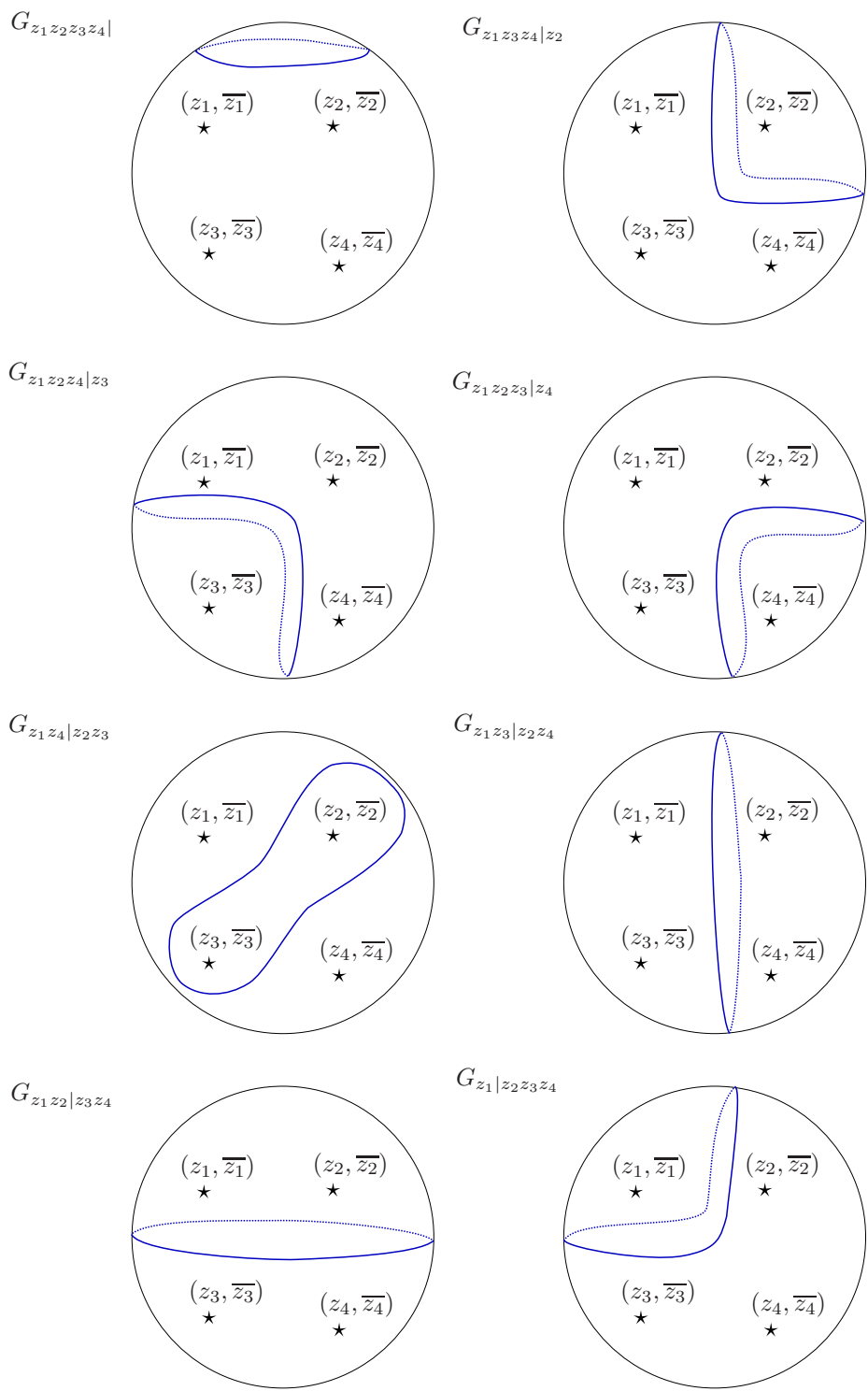}
\caption{The eight distinct configurations of a loop around a
given set of four points on the Riemann sphere.}\label{Fig4pts1loop}
\end{figure}
In this section, we expand the left hand side of equation~(\ref{Eq4ptInterpretation}) for $n\ll 1$ to order $n^1$.
The coefficient of $n^1$ may then be equated with the result for expansion of the right hand side,
described in the previous section and leading to equation~(\ref{Eq4ptSmallNCFT}).
Recall that the partition function for the loop gas expansion is equation~(\ref{EqLoopPartition}).
The sum may be decomposed into a term of order $n^0$ coming from the graph with no loops (hence the zeroth power
of $n$) and graphs with a single loop weighted by $n^1$. Other graphs will contribute terms of order $n^2$ or smaller.
The graphs with a single loop on the Riemann sphere may be further split up into eight distinct subsets,
which are shown in figure \ref{Fig4pts1loop}. Each configuration has a unique set of $(-1)^{N_{ij}}$
where $i\neq j$ and $i,j\in\{1,2,3,4\}$.

Let $L$ be defined as
$$
L\equiv\la(-1)^{N_{12}}(-1)^{N_{34}}\ra_{\textrm{loop gas}}
=\frac{\sum_{G}{(-1)^{N_{12}}(-1)^{N_{34}}n^{\textrm{\#loops}}x_{c}^{\textrm{total length of loops}}}}
{\sum_{G}{n^{\textrm{\#loops}}x_{c}^{\textrm{total length of loops}}}}\,.
$$
The numerator of $L$ can be calculated as
\begin{align*}
L_{\textrm{numerator}}&=1+n\Big[\sum_{G_{\z 1\z 2\z 3\z 4|}}{(1)(1)x_{c}^{l}}+
\sum_{G_{\z 1\z 2\z 3|\z 4}}{(1)(-1)x_{c}^{l}}
+\sum_{G_{\z 1\z 2\z 4|\z 3}}{(1)(-1)x_{c}^{l}}\\
&\qquad+\sum_{G_{\z 1\z 3\z 4|\z 2}}{(-1)(1)x_{c}^{l}}+\sum_{G_{\z 1\z 2|\z 3\z 4}}{(1)(1)x_{c}^{l}}
+\sum_{G_{\z 1\z 3|\z 2\z 4}}{(-1)(-1)x_{c}^{l}}\\
&\qquad+\sum_{G_{\z 1\z 4|\z 2\z 3}}{(-1)(-1)x_{c}^{l}}+\sum_{G_{\z 1|\z 2\z 3\z 4}}{(-1)(1)x_{c}^{l}}\
\Big]+O(n^2)\,.
\end{align*}
The denominator takes the same form as the expression above, but without the factors of $(1)$ and $(-1)$.
For ease of notation, define
$$
W_{\z 1\z 2\z 3\z 4|}=\sum_{G_{\z 1\z 2\z 3\z 4|}}{x_{c}^{l}}\,,
$$
and similar expressions for the other graphs in figure~\ref{Fig4pts1loop}.
The elements of $\{W_{i}\}$ are sums over all loops of a single configuration type $i$, weighted by $x_{c}$ to the
power of their length. They therefore represent the $\mu$-masses of loops of configuration $i$.
It may then be seen that
\begin{align}
L=1-2n\Big[W_{\z 1\z 2\z 3|\z 4}+W_{\z 1\z 2\z 4|\z 3}+W_{\z 1\z 3\z 4|\z 2}+W_{\z 1|\z 2\z 3\z 4}
\Big]+O(n^2)\label{Eq4ptCorrFnWExpansion}
\end{align}

The two point functions $C_{\{ij\}}\equiv\la \phi(\z i,\zb i)\phi(\z j,\zb j)\ra=\la (-1)^{N_{ij}}\ra$
and products of pairs of two point functions can be expanded in terms of the $\{W_{i}\}$ also:
\begin{align}
C_{12}C_{34}&=1-2n\big(W_{\z 1\z 3\z 4|\z 2}+W_{\z 1\z 2\z 4|\z 3}+W_{\z 1\z 2\z 3|\z 4}\nonumber\\
&\qquad+W_{\z 1|\z 2\z 3\z 4}+2W_{\z 1\z 4|\z 2\z 3}+2W_{\z 1\z 3|\z 2\z 4}\big)+O(n^2)\label{EqC12C34WExpansion}\\
C_{13}C_{24}&=1-2n\big(W_{\z 1\z 3\z 4|\z 2}+W_{\z 1\z 2\z 4|\z 3}+W_{\z 1\z 2\z 3|\z 4}\nonumber\\
&\qquad+W_{\z 1|\z 2\z 3\z 4}+2W_{\z 1\z 4|\z 2\z 3}+2W_{\z 1\z 2|\z 3\z 4}\big)+O(n^2)\label{EqC13C24WExpansion}\\
C_{14}C_{23}&=1-2n\big(W_{\z 1\z 3\z 4|\z 2}+W_{\z 1\z 2\z 4|\z 3}+W_{\z 1\z 2\z 3|\z 4}\nonumber\\
&\qquad+W_{\z 1|\z 2\z 3\z 4}+2W_{\z 1\z 2|\z 3\z 4}+2W_{\z 1\z 3|\z 2\z 4}\big)+O(n^2)\label{EqC14C23WExpansion}\,.
\end{align}
Thus, it can be seen from
equations~(\ref{Eq4ptCorrFnWExpansion}),(\ref{EqC12C34WExpansion}),(\ref{EqC13C24WExpansion})~and~(\ref{EqC14C23WExpansion})
that a subset $\{W_{c}\}$ of the $\{W_{i}\}$ can be expressed in terms of
partly-connected four point functions:
\begin{align}
W_{\z 1\z 4|\z 2\z 3}&=\frac{1}{8n}(L-C_{12}C_{34}-C_{13}C_{24}+C_{14}C_{23})+O(n)\label{EqW14FnLCs}\\
W_{\z 1\z 3|\z 2\z 4}&=\frac{1}{8n}(L-C_{12}C_{34}+C_{13}C_{24}-C_{14}C_{23})+O(n)\\
W_{\z 1\z 2|\z 3\z 4}&=\frac{1}{8n}(L+C_{12}C_{34}-C_{13}C_{24}-C_{14}C_{23})+O(n)\,.
\end{align}
These $\{W_{c}\}$ correspond to graphs with a loop winding around two of the four points
and hence do not include contributions from vanishingly small loops in the continuum limit.
By comparison with the small $n$ expansions of $L$ (section~\ref{Sec4ptSmallN}) and
$C_{\{ij\}}$ (section \ref{Sec2ptsmalln} with the same choice for the normalisation of the twist operator
as in the four point function) from conformal field theory, these weights
are the following functions of $\eta,\overline{\eta}$
\begin{align}
W_{\z 1\z 4|\z 2\z 3}&=\frac{-1}{6\pi}\ln|1-\eta|+q(\eta,\overline{\eta})
+O(n)\label{EqW14CFT}\\
W_{\z 1\z 3|\z 2\z 4}&=q(\eta,\overline{\eta})
+O(n)\label{EqW24CFT}\\
W_{\z 1\z 2|\z 3\z 4}&=\frac{-1}{6\pi}\ln|\eta|+q(\eta,\overline{\eta})
+O(n)\label{EqW34CFT}\,,
\end{align}
where
\begin{align*}
q(\eta,\overline{\eta})=
\frac{-1}{24\pi}&\Big(\eta\,_{3}F_{2}(1,1,\frac{4}{3};2,\frac{5}{3};\eta)
+\overline{\eta}\,_{3}F_{2}(1,1,\frac{4}{3};2,\frac{5}{3};\overline{\eta})\Big)\\
&+\frac{2^{1/3}\pi}{3\sqrt{3}\Gamma(\frac{1}{6})^2\Gamma(\frac{4}{3})^2}
|\eta(1-\eta)|^{2/3}|\,_{2}F_{1}(\frac{2}{3},1,\frac{4}{3},\eta)|^{2}\,.
\end{align*}
Note that the $W_{c}$ above are finite in the continuum limit of vanishing lattice spacing.
In fact, the expressions in equations~(\ref{EqW14CFT}),(\ref{EqW24CFT})~and~(\ref{EqW34CFT}) are independent of $a$.
They are finite and non-zero in the limit $n\rightarrow 0$ and are invariant under
conformal transformations, being functions only of the cross ratios.
These expressions for the $\mu$-masses $\{W_{c}\}$ constitute one of the main results of this paper.
\subsection{The central charge}
It is a standard result of conformal field theory that, given the
explicit form of any four-point function, the central charge $c$
may be determined. This is because the operator product expansion
(OPE) of any scalar primary operator $\phi$ with itself takes the
form
\begin{equation}
\phi(z_1,\bar z_1)\phi(z_2,\bar
z_2)=|z_{12}|^{-4h}\left(1+\cdots+\frac{2h}{c}T(z_1)+\cdots\right)
\end{equation}
where $T$ is the holomorphic component of the stress tensor. The
coefficient follows from consideration of the limit $z_{12}\to0$
in the three point function $\langle
T(z)\phi(z_1)\phi(z_2)\rangle\propto 2h$ and the two-point
function $\langle T(z)T(z_1)\rangle=(c/2)(z-z_1)^{-4}$.

If this result is applied to the explicit expression (\ref{Eq4ptCFTResult}),
we find that the coefficient of $\eta^2$ is
$$
h_{2}(2h_{2}+1)+\frac{2h_{2}(1-\frac{\kappa}{4})(2-\frac{3\kappa}{4})}{2-\frac{\kappa}{2}}
+\frac{(1-\frac{\kappa}{4})(2-\frac{\kappa}{4})(2-\frac{3\kappa}{4})(3-\frac{3\kappa}{4})}
{2(2-\frac{\kappa}{2})(2-\frac{\kappa}{2})}\,,
$$
and so obtain the known expression for the central charge in terms of
$\kappa$
\begin{equation}
c=\frac{(3\kappa-8)(6-\kappa)}{2\kappa}\,.
\end{equation}

This result is of interest in the limit $n\to0$ for the
light it sheds on the interpretation of the stress tensor $T$ in
terms of an observable of a random curve given by Doyon, Riva and
Cardy \cite{DoyonRivaCardy}. In that paper it was shown that, for any measure
on simple random curves which satisfies conformal restriction, one
may identify $T(z)$ as being proportional to the spin-2 angular
Fourier component of the probability that the curve intersects a
small line segment of length $\epsilon$ centred at $z$. While in
that paper the focus was on curves which connect two points on
the boundary of a simply connected domain, described in the case
when conformal restriction holds by SLE$_{8/3}$, the theorem
should equally well apply to Werner's measure on self-avoiding
loops, if suitably re-interpreted in terms of $\mu$-masses rather
than probabilities.
However our results in the present paper do not directly apply to
the intersection with a line segment, but rather to the event that the
loop passes (or does not pass) between pairs of points (which, however, may
be taken to mark the ends of the line segment). Nevertheless, one
might expect these to differ only by a constant of
proportionality, and indeed our results in this paper support this,
and suggest that the two-point function of the object $T$ introduced
in Ref.~\cite{DoyonRivaCardy} is indeed given by the central charge for small $n$ as
expected.

Indeed, let us consider the quantity $W_{z_1z_4|z_2z_3}$ as given
in equation~\ref{EqW14CFT}. This is the $\mu$-mass of loops which separate
$(z_1,z_4)$ from $(z_2,z_3)$. In the limit $z_{12}\to0$, writing
$z_{12}=\epsilon e^{i\theta_{12}}$, we can define
\begin{equation}
V(z_1;z_3,z_4)=5\lim_{\epsilon\to0}\epsilon^{-2}\int
d\theta_{12}e^{-2i\theta_{12}}W_{z_1z_4|z_2z_3}\,,
\end{equation}
where the numerical prefactor is $\lim_{n\to0}(c/2h_2)$. Our
result equation~\ref{Eq4ptCFTResult} then implies that $V$ has the
same dependence on its arguments as does the O$(n)$ term in the
CFT correlation function
\begin{equation}
\langle T(z_1)\phi_{1,2}(z_3,\bar z_3)\phi_{1,2}(z_4,\bar
z_4)\rangle\,.
\end{equation}
This result generalises to other correlation functions, and
implies that we may interpret the spin-2 component of the
$\mu$-mass of loops which pass between two nearby points as being,
in some sense, the derivative $\widetilde T$ of the CFT stress
tensor with respect to $n$ at $n=0$. With this definition we then
find the result
\begin{equation}
\langle\widetilde T(z_1)\widetilde
T(z_3)\rangle=\frac{c'(0)/2}{(z_1-z_3)^4}\,,
\end{equation}
as expected, where the left hand side is
\begin{equation}
5^2\lim_{\epsilon\to0}\epsilon^{-4} \int
d\theta_{12}e^{-2i\theta_{12}} \int
d\theta_{34}e^{-2i\theta_{34}}W_{z_1z_4|z_2z_3}
\end{equation}

It would, of course, be important to establish this interpretation
directly from the restriction property.
\section{Twist operators in a simply connected domain}\label{Sec2ptBCFT}
The model may also be considered in a simply connected domain.
All such domains may be conformally transformed to the upper half plane, with the real axis
being the boundary.
We may therefore derive the results for the upper half plane.
The two point function of twist operators in the presence of this boundary satisfies the same set of partial
differential equations
as the four point function in the bulk, with $(\z 3,\zb 3)$,$(\z 4,\zb 4)$ assigned as the complex conjugates
of the positions of the operators. This is a well known result in the theory of boundary conformal field
theory~\cite{bigyellow} (BCFT) and follows from the condition of $T=\overline{T}$ on the boundary.
The solution for the two point function is
\begin{align}
\la&\phi_{2}(\z 1)\phi_{2}(\z 2)\ra_{\textrm{BCFT}}=\Big(\frac{(\z 1-z_{1}^{*})(\z 2-z_{2}^{*})a^{2}}
{(\z 1-\z{2})(z_{1}^{*}-z_{2}^{*})(\z 2-z_{1}^{*})(\z 1-z_{2}^{*})}\Big)^{3\kappa/8-1}\times\nonumber\\
&A(\kappa)\Big[
_{2}F_{1}(1-\frac{\kappa}{4},2-\frac{3\kappa}{4};2-\frac{\kappa}{2};\eta)
+B(\kappa)(-\eta(1-\eta))^{\kappa/2-1}\,_{2}F_{1}(\frac{\kappa}{4},\frac{3\kappa}{4}-1;\frac{\kappa}{2};\eta)
\Big]\label{EqBCFTResult}\,,
\end{align}
where now there is only one cross-ratio
$$
\eta=\frac{(z_{1}-z_{2})(z_{1}^{*}-z_{2}^{*})}{(z_{1}-z_{1}^{*})(z_{2}-z_{2}^{*})}\,.
$$
$B(\kappa)$ can be determined by looking at the boundary conditions as the two points approach the
boundary, which is
$\eta\rightarrow -\infty$. Equation~(\ref{EqBCFTResult}) may be analytically continued to large $\eta$:
\begin{align}
&\Big(\frac{(z_{2}-z_{1}^{*})(z_{2}^{*}-z_{1})}{a^{2}}\Big)^{1-\frac{3\kappa}{8}}A(\kappa)\times\\
&\bigg[
\big(\frac{-1}{\eta}\big)^{\frac{\kappa}{8}}\,_{2}F_{1}\big(1-\frac{\kappa}{4},\frac{\kappa}{4};\frac{\kappa}{2}
;\frac{1}{\eta}\big)\Big(
\frac{\Gamma(2-\frac{\kappa}{2})\Gamma(1-\frac{\kappa}{2})}{\Gamma(2-\frac{3\kappa}{4})
\Gamma(1-\frac{\kappa}{4})}+B\frac{\Gamma(\frac{\kappa}{2})\Gamma(1-\frac{\kappa}{2})}
{\Gamma(\frac{\kappa}{4})\Gamma(1-\frac{\kappa}{4})}\Big)+\nonumber\\
&(-\eta)^{3\kappa/8-1}\,_{2}F_{1}\big(2-\frac{3\kappa}{4},\frac{4-\kappa}{4};2-\frac{\kappa}{2};\frac{1}{\eta}\big)
\Big(\frac{\Gamma(2-\frac{\kappa}{2})\Gamma(\frac{\kappa}{2}-1)}{\Gamma(1-\frac{\kappa}{4})\Gamma(\frac{\kappa}{4})}
+B\frac{\Gamma(\frac{\kappa}{2})\Gamma(\frac{\kappa}{2}-1)}{\Gamma(\frac{3\kappa}{4}-1)\Gamma(\frac{\kappa}{4})}\Big)
\bigg]\label{EqBCFTLargeEta}\,.
\end{align}
The two natural choices of $B(\kappa)$ are those which pick out one or the other conformal block as the operators approach
the boundary.
In the limit $n\rightarrow 0$, there are no loops in the loop gas picture so the correlation function should tend to
unity for all $\eta$.
This is only possible if the first term in the square brackets of equation~(\ref{EqBCFTLargeEta})
is not present. A (non-unique) choice of $B(\kappa)$ which satisfies this requirement is
that which picks out the second conformal block only:
$$
B(\kappa)=-\frac{\Gamma(2-\frac{\kappa}{2})\Gamma(1-\frac{\kappa}{2})\Gamma(\frac{\kappa}{4})\Gamma(1-\frac{\kappa}{4})}
{\Gamma(\frac{\kappa}{2})\Gamma(1-\frac{\kappa}{2})\Gamma(2-\frac{3\kappa}{4})\Gamma(1-\frac{\kappa}{4})}\,.
$$
It is important to note also that this boundary condition is not compatible with
the vanishing of the four point function in the limit $\eta\rightarrow -\infty$.
\begin{figure}[t]
\centering
\includegraphics[width=1.1\textwidth]{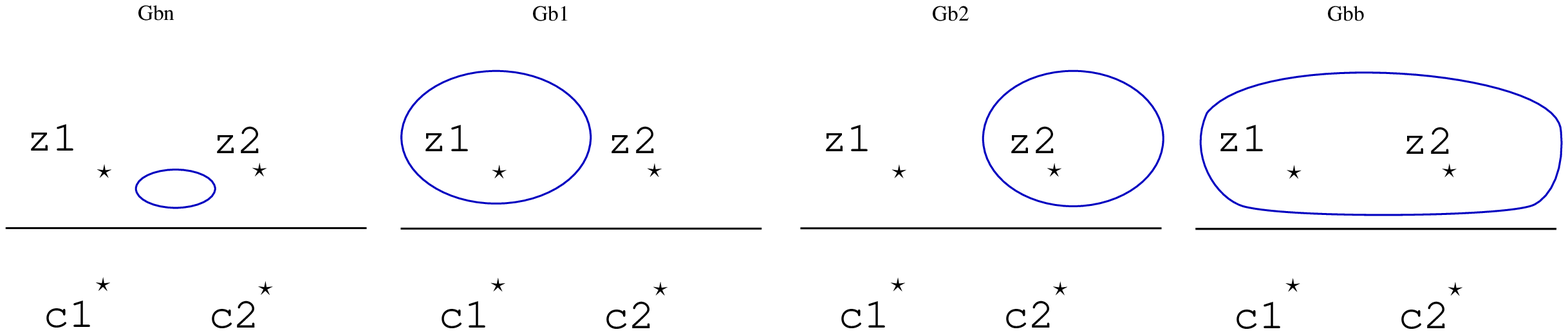}
\caption{The four types of configuration of a single loop around two points in the presence of
a boundary. The complex conjugate points are included only for ease of calculation of
quantities such as $(-1)^{N_{1,1*}}$ (see text).}\label{Fig2ptBoundary1loop}
\end{figure}
The form of $A(\kappa)$ is dependent on the choice of normalisation of the twist operators.
This can be seen from the $\eta\rightarrow 0$ limit of equation~(\ref{EqBCFTResult}):
$$
\lim_{\eta\rightarrow 0}{\la \phi_{2}(z_{1})\phi_{2}(z_{2})\ra}=A(\kappa)\Big(
\frac{(z_{1}-z_{2})(z_{1}^{*}-z_{2}^{*})}{a^2}\Big)^{1-3\kappa/8}\,.
$$
$A(\kappa)$ must be of the form $A=1+\sigma n+O(n^2)$
due to the constraint that the four point function tends to unity as $n\rightarrow 0$.
For small $n$, the correlation function may be expanded as
\begin{align*}
\la\phi_{2}(\z 1)\phi_{2}(\z 2)&\ra_{\textrm{BCFT}}=
1+n\Big[\frac{1}{3\pi}\ln(\mu)-\frac{1}{3\pi}\eta\,_{3}F_{2}(1,1,\frac{4}{3};2,\frac{5}{3};\eta)\\
&+\frac{2\Gamma(\frac{2}{3})^{2}}{3\pi\Gamma(\frac{4}{3})}(-\eta(1-\eta))^{\frac{1}{3}}
\,_{2}F_{1}(\frac{2}{3},1;\frac{4}{3};\eta)+\sigma\Big]+O(n^2)\,,
\end{align*}
where
$$
\mu=\frac{(\z 1-z_{1}^{*})(\z 2-z_{2}^{*})a^2}
{(\z 1-\z{2})(z_{1}^{*}-z_{2}^{*})(\z 2-z_{1}^{*})(\z 1-z_{2}^{*})}\,.
$$
In analogy with the interpretation of the four point function in the bulk in the loop gas picture
described in section~\ref{Sec4ptInterpretation}, the following is the interpretation of the
two point function in the presence of a boundary
\begin{equation}\label{Eq2ptBoundaryInterpretation}
\la (-1)^{N_{1,1^{*}}}(-1)^{N_{2,2^{*}}}\ra_{\textrm{loop gas}}=\la\phi(\z 1,\zb 1)\phi(\z 2,\zb 2)\ra_{\textrm{BCFT}}\,.
\end{equation}
The left hand side is an expectation value in the ensemble of lattice loops, as before.
$N_{1,1^{*}}$ is the number of times loops cross a defect line from $z_{1}$ and $z_{1}^{*}$.
The set of all graphs may be decomposed into the set with no loops and the set with one loop;
the other graphs are of order $n^2$.
The possible configurations of a single loop with a boundary are shown in figure~\ref{Fig2ptBoundary1loop}.
The two and one point functions are then found from the loop gas to be
\begin{align*}
M&\equiv\la\phi(\z 1)\phi(\z 2)\ra=\la (-1)^{N_{1,1^{*}}}(-1)^{N_{2,2^{*}}}\ra
=1-2n(W_{1(2)}+W_{(1)2})+O(n^2)\\
C_{1}&\equiv\la\phi(z_{1})\ra=\la (-1)^{N_{1,1*}}\ra
=1-2n(W_{(1)2}+W_{(12)})+O(n^2)\\
C_{2}&\equiv\la\phi(z_{2})\ra=\la (-1)^{N_{2,2*}}\ra
=1-2n(W_{1(2)}+W_{(12)})+O(n^2)\,.
\end{align*}
Just as for the case of the loop gas in the bulk,
the $\mu$-masses $\{W_{i}\}$ are the number of weighted loops belonging
to the configuration $i$ (see figure~\ref{Fig2ptBoundary1loop}).
They are defined as
$$
W_{1(2)}\equiv\sum_{G_{1(2)}}{x_{c}^{l}}\,.
$$
Of the four possible configurations, the only one finite in the limit $a\rightarrow 0$ is $W_{(12)}$:
$$
W_{(12)}=\frac{M-C_{1}C_{2}}{4n}+O(n)\,.
$$
In terms of $\eta$,
\begin{align}
W_{(12)}=-\frac{1}{12\pi}&\ln(\eta(1-\eta))-\frac{1}{12\pi}\eta\,_{3}F_{2}(1,1,\frac{4}{3};2,\frac{5}{3};\eta)\nonumber\\
&+\frac{\Gamma(2/3)^2}{6\pi\Gamma(4/3)}(-\eta(1-\eta))^{\frac{1}{3}}\,_{2}
F_{1}(\frac{2}{3},1;\frac{4}{3};\eta)+O(n)\label{Eq2ptBoundaryScaleInvW}\,.
\end{align}
Note, as for the bulk case, that the term involving $\sigma$ in $M$ is cancelled by the
subtraction of the product of one point functions.
\section{Interpretation as a stochastic process}\label{SecStochasticProcess}
We have seen in section \ref{Sec4ptInterpretation} that the
$\mu$-mass of loops which wind around two of four points is
invariant under conformal transformations. In this section, we
show that the differential equations they satisfy can be
interpreted in terms of an SLE process. Recall that these
functions were identified as semi-connected four point
functions, for example equation~(\ref{EqW14FnLCs})
\begin{align*}
W_{\z 1\z 4|\z 2\z 3 }=\lim_{n\to0}(8n)^{-1}
\Big[\la\phi(z_{1},&\overline{z_{1}})\phi(z_{2},\overline{z_{2}})\phi(z_{3},\overline{z_{3}})\phi(z_{4}
,\overline{z_{4}})\ra\\
&-C_{12}C_{34}-C_{13}C_{24}+C_{14}C_{23}\Big]\,,
\end{align*}
where $C_{\{ij\}}$ are the correlation functions of a pair of
twist operators at the points $(\z i,\zb i)$ and $(\z j,\zb j)$.
The two point function is fixed by scale invariance and the four
point function was determined as a solution to the partial
differential
equations~(\ref{Eq4ptBPZHolomorphic})~and~(\ref{Eq4ptBPZAntiholomorphic}).
Using these, we find that $W_{\z 1\z 4|\z 2\z 3}$ satisfies the
PDE
\begin{align}
\Big[\frac{3}{2}&\partial_{\z 1}^{2}+\sum_{i\neq
1}\frac{\partial_{\z i}}{z_{i}-\z 1} \Big]W_{\z 1\z 4|\z 2\z
3}=\frac{1}{24\pi}\Big[
\frac{1}{(\z 4-\z 1)^2}+\frac{1}{\z 3-\z 1}\Big(\frac{1}{\z 3-\z 4}+\frac{1}{\z 2-\z 3}\Big)\nonumber\\
&+\frac{1}{\z 2-\z 1}\Big(\frac{1}{\z 2-\z 4}+\frac{1}{\z 3-\z
2}\Big) +\frac{1}{\z 4-\z 1}\Big(\frac{1}{\z 4-\z 3}+\frac{1}{\z
4-\z 2}\Big) \Big]+O(n)\label{EqnWholo}\,,
\end{align}
together with a similar equation in which all the $z\,$s are
replaced by $\bar z$. Note that in the BPZ equations $z$ and $\bar
z$ can be taken as independent complex numbers, and it is only in
applying these equations to physical quantities that one needs to
impose reality conditions. These equations have almost the second
order linear form which would result from applying the It\^{o} formula
to a martingale of an SLE process started at $z_1$. One difference
is that they are complex. We can obtain a real equation by, for
example, taking the real part of the sum of the holomorphic and
antiholomorphic equations (the more general case will be discussed
below), but note that when we do this we have the sum
$$
\partial_{z_1}^2+\partial_{\bar z_1}^2=
(\partial_{z_1}+\partial_{\bar
z_1})^2-2\partial_{z_1}\partial_{\bar z_1} =
\partial_{x_1}^2-\frac12\Delta_{z_1}\,,
$$
where $x_1=\Re e(z_1)$ and $\Delta_{z_1}$ is the Laplacian operator
$\partial_{x_1}^2+\partial_{y_1}^2$. We also note that we can
remove the inhomogenous part in (\ref{EqnWholo}) by defining
\begin{equation}
\label{Eqnsubtract} \widetilde W\equiv W-(24\pi)^{-1}2\Re e \big[
-2\ln(z_4-z_1)+\ln(z_3-z_4)+\ln(z_2-z_4) -\ln(z_2-z_3) \big]
\end{equation}
Then we can rewrite the real part of (\ref{EqnWholo}) as
\begin{align}
\Big[3\partial_{x_1}^{2}&+2\Re e\sum_{i\neq 1}\frac{2\partial_{\z
i}}{z_{i}-\z 1} \Big]\widetilde W=\frac32\Delta_{z_1} \widetilde
W\,.\label{EqWPDE}
\end{align}

Consider now a sequence of conformal maps $g_t(z)$ which satisfy
the stochastic chordal Loewner equation
\begin{align}
\label{CSLE} \frac{dg_t(z)}{dt}=\frac{2}{g_t(z)-z_{1t}}\,,
\end{align}
where $z_{1t}=z_{1}+\sqrt\kappa B_{t}$.
Then if $M\big(z_{1t},g_t(z_j)\big)$ were a
martingale, its expectation value would satisfy (\ref{EqWPDE})
with the right-hand side set equal to zero. The coefficient of
$\partial_{x_1}^2$ would in general be $\kappa/2$, so we should
take $\kappa=6$. The right-hand side of (\ref{EqWPDE}), together with the subtractions in~(\ref{Eqnsubtract}),
therefore express the degree to which the weighted number of loops separating
$(z_1,z_4)$ from $(z_2,z_3)$ fails to be a martingale.

\begin{figure}[t]
\centering
\includegraphics[width=0.8\textwidth]{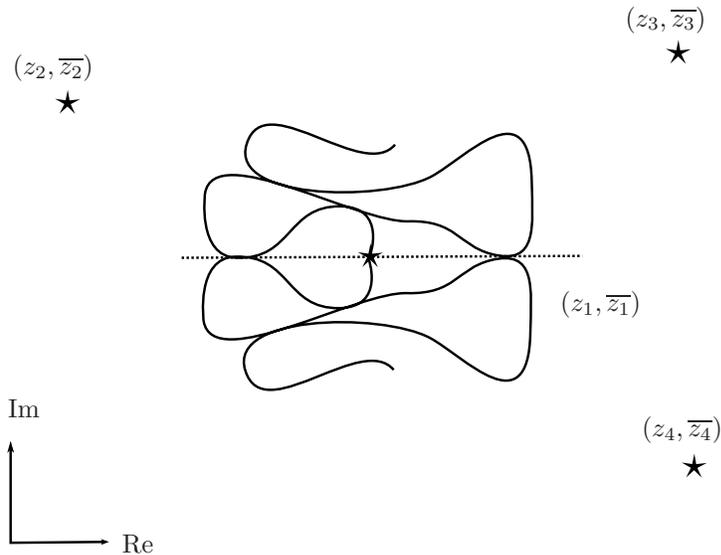}
\caption{A chordal SLE$_{8/3}$ growing in the half-plane $\Im m(z)>\Im m(z_1)$ with its reflection in
$\Im m(z)=\Im m(z_1)$}\label{FigSLE}
\end{figure}

Eq.~\ref{CSLE} describes a chordal SLE growing into the half-plane
$\Im m(z)>\Im m(z_1)$, together with its exact reflection in the line
$\Im m(z)=\Im m(z_1)$ as in figure~\ref{FigSLE}~\cite{CardyRev}. For $\kappa>4$ the hull, or outer perimeter, of
this curve encloses a growing region of the plane. The entire
interior of this hull at time $t$ is mapped into the point
$\sqrt\kappa B_t$. Because of this, we would not expect the
weighted number of loops to be a martingale. Given some initial
population of loops, each time that part of a loop is swallowed by
the growing hull, it disappears from the population. The
right-hand side of (\ref{EqWPDE}) must therefore express the rate
at which this happens. It is reasonable that this should be
proportional to $\Delta_{z_1} W$: if $W$ were constant in some region,
this would mean there were no loops passing through it. The
leading non-zero rotationally invariant contribution should therefore be
proportional to the Laplacian.

Apart from this, the left-hand side expresses the fact that the
$\mu$-mass of loops which separate $z_1z_4$ and $z_2z_3$ and which
have not yet been (partly or wholly) swallowed by the hull is the
same as that in the conformally equivalent case of loops which
separate $(z_{1t},g_t(z_4))$ from $(g_t(z_2),g_t(z_3))$.
This breaks down when the hull swallows any of the other three points.
We conjecture that the subtractions in~(\ref{Eqnsubtract}) take this into account.

Of course, this is only suggestive and a number of important
issues would have to be resolved before one could actually derive our
results from SLE. In particular, one should explain why it is
necessary to consider SLE$_6$ rather than some other value of
$\kappa$. This is presumably related to the fact that the hull of
SLE$_6$ corresponds locally to SLE$_{8/3}$, and that both correspond
to CFTs with central charge $c=0$. The choice of a chordal SLE reflected in
$\Im m(z)=\Im m(z_1)$ is clearly arbitrary; other choices correspond to taking
different linear combinations of the holomorphic and anti-holomorphic equations.
Perhaps using radial or whole-plane SLE would make the formula look more symmetrical.
\section{Conclusion}
In this paper we have studied scaling properties of loops in the loop gas picture of the O($n$) model.
In the picture, the partition function is a sum over all graphs of non-intersecting closed loops, weighted by $nx_{c}^{l}$
where $x$ is a function of the reduced coupling and $l$ is the length of the loop. We introduced twist operators, whose
correlation functions count the loops separating
the locations of the operators with weight $n'$ different to the usual weight $n$, or equivalently
count the minimum number of crossings of defect lines running between these locations.
For the particular choice $n'=-n$, the twist operators have level two null states and their correlation functions satisfy
BPZ type partial differential equations on the Riemann sphere.
Thus, conformal field theory may be used to determine the analytic form of the two and four point functions.
In the loop gas picture, the choice $n'=-n$ means that loops are counted are weighted by an additional
factor of $-1$ to the
power of the number of defect lines crossed, and the choice of path for the defect lines
between the locations of the operators is unimportant.

The limit $n\rightarrow 0$ describes the theory of self-avoiding loops.
In this limit,
the partition function and correlation functions are dependent to first order in~$n$ only on the
configurations of a single loop. Hence, by equating particular semi-connected parts of the four point function
calculated using conformal field theory
to the result from the Coulomb gas picture, we have deduced the expected number of weighted loops winding around two of the
four points and shown that this number is invariant under conformal transformations.
This is presumably the mass of such a subset under the measure on loops introduced by Werner.
Other configurations of loops receive
contributions from vanishingly small loops around a single point and are not finite in the limit of vanishing
lattice spacing.

The central charge of the O($n$) model for small $n$ may be found
from the dependence of the $\mu$-mass of loops around two of the
four points as a function of the cross ratio of the positions of
the four operators. The result $c\sim 5n/3\pi$ from the analytic
results agrees with other methods of calculating the central
charge of the O($n$) model, and lends support to the
interpretation of the stress tensor for curves satisfying
conformal restriction given in Ref.\cite{DoyonRivaCardy}.

A similar calculation was also carried out for the model defined in a simply connected domain.
A general simply connected domain may be mapped
via a conformal transformation to the upper half plane with the real axis as the boundary. The two point
function of operators in the upper half plane satisfies the same partial differential equations as the
four point function on the Riemann sphere,
with two additional operators positioned at the complex conjugates of the original operators.
By again equating the results from conformal field theory with those from the Coulomb gas picture, we have
deduced the $\mu$-mass of loops around the two points in the upper half plane.

Finally, we have shown that the differential equations satisfied by the above quantities should
have a stochastic interpretation in terms of a chordal SLE$_{6}$ process
starting from one of the points $z_{1}$, along with its reflection in a fixed line.

\em Acknowledgments\em:

We thank Benjamin Doyon and Valentina Riva for useful discussions.
This work was supported in part by EPSRC Grant GR/R83712/1. AG was
supported by an EPSRC Studentship.

\end{document}